# Car-following model on two lanes and stability analysis[*]

Jia Yu-han(贾宇涵)[a),†], Wu Jian-ping(吴建平)[a)], Du Yi-man(杜怡曼)[a)], Qi Ge-qi(奇格奇)[a)]

[a] Department of Civil Engineering, Tsinghua University, 100084, Beijing, China

**Abstract:** Considering lateral influence from adjacent lane, an improved car-following model is developed in this paper. Then linear and non-linear stability analyses are carried out. The modified Korteweg-de Vries (MKdV) equation is derived with the kink-antikink soliton solution. Numerical simulations are implemented and the result shows good consistency with theoretical study.

## 1. Introduction

Traffic flow is a system of consecutive vehicles with interacting [1]. Recently various models have been developed, including general models, safety distance models, action point models, optimal velocity models (OVM), cellular automaton models and fuzzy logic models [2-5]. Among those models, OVM developed by Bando *et al*. [6,7] is well known for its accuracy and rationality. Afterwards Helbing and Tilch [8] calibrated the OV model by experimental data and developed a generalized force model (GFM) to overcome the deficiencies. But both OVM and GFM cannot describe the phenomenon that the following vehicle may not decelerate when the leading vehicle is much faster even if the headway distance is smaller than safety distance. After this, a full velocity difference model (FVDM) was developed by Jiang *et al*. [9,10] to solve the disadvantage.

Based on OVM, GFM and FVDM, many new models have been established by considering decentralized delayed-feedback control [11], delay time due to driver's reaction [12], extended OV function for cooperative driving control system [13,14], acceleration difference [15], multiple velocity difference [16], optimal velocity difference [17], and control method [18].

To study traffic jam waves in OVM, Komatsu Sasa [19] firstly derived the modified Korteweg-de Vries (MKdV) equation to describe kink waves. Then Muramatsu and Nagatani [20] derived Korteweg-de Vries (KdV) equation from OVM to describe sliton waves in traffic jam, and Nagatani also found triangular shock wave solved Burgers equation [21]. From then many models have been analyzed by non-linear stability theory aforementioned. Nagatani [22] derived MKdV equation near critical point in two continuum models: partial differential and discrete lattice model. Xue et al. [23,24] presented a simplified OVM considering relative velocity and derived KdV and MKdV equations. Ge et al. developed several intelligent transportation system (ITS) based models with KdV and MKdV analysis [25,26] and also did similar research in three OVM based models [27]. Yu [28] recently build a two-delay model with MKdV investigation and implemented numerical simulations. More studies show that the triangular wave, soliton wave and kink wave occur in stable region, metastable region and unstable region, respectively [21,29,30].

[*] Project funded by the National High Technology Research and Development Program of China (Grant NO. 2012AA063303)
[†] Corresponding Author: jiajiayuyuhanhan@163.com

However, only a few researches focused on car-following with lateral impact, in which case the lateral influence from adjacent lane should be considered. Nagatani [31] presented two lattice models to simulate traffic flow wave on a two-lane highway with lane changing. Tang et al. [32] studied the stability of a two-lane OVM based model with MKdV analysis by defining a weighted headway distance. Peng [33]. Jin et al. [34] considered the lane-width influence and developed a non-lane-based FVDM with simulation experiments. Ge et al. [35,36] studied the influence from neighbor vehicle or non-motor vehicle by considering two more OV functions and analyzed the stability condition by control theory method. Based on previous work, this paper investigates a new car-following model considering lateral influence by introducing the combination of two OV functions. In section 2 the new model is developed and linear stability analysis is carried out in section 3. In section 4 the MKdV equation is derived to obtain kink-antikink soliton solution. Then numerical simulation experiments are performed to verify the theoretical study in section 5. The summary is given in section 6.

## 2. Improved OVM

The typical OVM is presented as [6,7]

$$\frac{d^2 x_n(t)}{dt^2} = \alpha [V^{op}(\Delta x_n(t)) - v_n(t)], \quad (1)$$

where $x_n(t)$ and $v_n(t)$ are the position and velocity of the $n$th vehicle; $\Delta x_n(t)$ is the headway distance between the $n$th and its leading vehicle; $\alpha$ is the sensitivity parameter of the driver; $V^{op}(\cdot)$ is the optimal velocity function described as [6]

$$V^{op}(\Delta x_n(t)) = \frac{v_{\max}}{2}[\tanh(\Delta x_n(t) - h_c) + \tanh(h_c)], \quad (2)$$

where $v_{\max}$ is the maximum velocity on a particular roadway; $h_c$ means the safety headway distance.

However, as noticed in the study on roadway, a driver usually focuses not only the lead-ing vehicle on the present lane, but also the vehicle on adjacent lane, especially when the neighbor vehicle decelerates. This phenomenon occurs because of the potential action of lane changing or the avoidance of collision when the lane width is small [32]. Hence the lateral influence should be considered in car-following model even if lane changing does not occur.

It is assumed that the driver makes his decision upon the combination impact of leading vehicle and neighbor vehicle by introducing a second OV function, which can be defined as

$$\bar{V}^{op}(l_n(t)) = \begin{cases} V^{op}(\Delta x_{l,n+1}(t)), & l_v \leq \Delta x_{l,n+1}(t) < d \\ 0, & others \end{cases}, \quad (3)$$

where $\Delta x_{l,n+1}(t)$ is the headway distance between the *n*th vehicle and its leading vehicle on the adjacent lane, $l_v$ is the length of a normal vehicle and $d$ is a preset constant.

Referring to previous study [8,9], $\Delta v_n(t)$ and $\Delta v_{l,n}(t)$ are introduced, where $\Delta v_n(t) = v_{n+1}(t) - v_n(t)$ and $\Delta v_{l,n}(t)$ can be given as

$$\Delta v_{l,n}(t) = \begin{cases} v_{l,n}(t) - v_n(t), & l_v \leq \Delta x_{l,n+1}(t) < d \\ 0, & others \end{cases}, \quad (4)$$

in which $v_{l,n}(t)$ is the velocity of the leading vehicle on adjacent lane.

The new model can be expressed as

$$\frac{d^2 x_n(t)}{dt^2} = \alpha[pV^{op}(\Delta x_n(t)) + q\bar{V}^{op}(\Delta x_{l,n}(t)) \\ - v_n(t)] + \lambda_1 \Delta v_n(t) + \lambda_2 \Delta v_{l,n}(t), \quad (5)$$

where $p$ and $q$ are the weights of the two OV functions; $\lambda_1$ and $\lambda_2$ are the weights of velocity difference.

## 3. Linear stability analysis

According to linear stability analysis method [7,16], stable condition of the uniform traffic flow is given by

$$\begin{cases} x_n^{(0)}(t) = nh_c + V^{op}(h)t \\ x_{l,n}^{(0)}(t) = nh_c + \bar{V}^{op}(h)t \end{cases}. \quad (6)$$

Let $y_n(t)$ and $y_{l,n}(t)$ be small deviations from $x_n^{(0)}(t)$ and $x_{l,n}^{(0)}(t)$ as $x_n(t) = x^{(0)}(t) + y_n(t)$ and $x_{l,n}(t) = x_l^{(0)}(t) + y_{l,n}(t)$. The linearized equation can be obtained:

$$\frac{d^2 y_n(t)}{dt^2} = \alpha[pV'(\Delta x^{(0)})\Delta y(t) + q\bar{V}'(\Delta x_l^{(0)})\Delta y_l(t) \\ - v_n(t)] + \lambda_1 \Delta v_n(t) + \lambda_2 \Delta v_{ln}(t), \quad (7)$$

where $V'$ and $\bar{V}'$ are the derivatives of OV function $V^{op}(\Delta x_n(t))$ and $\bar{V}^{op}(l_n(t))$. Expanding $y_n(t) \propto \exp(ikn + zt)$ and $y_{l,n}(t) \propto \exp(ikn + zt)$, Eq. (7) can be rewritten as

$$z^2 + [a - (\lambda_1 + \lambda_2)(e^{ik} - 1)]z - \\ a(pV' + q\bar{V}')(e^{ik} - 1) = 0 \quad . \quad (8)$$

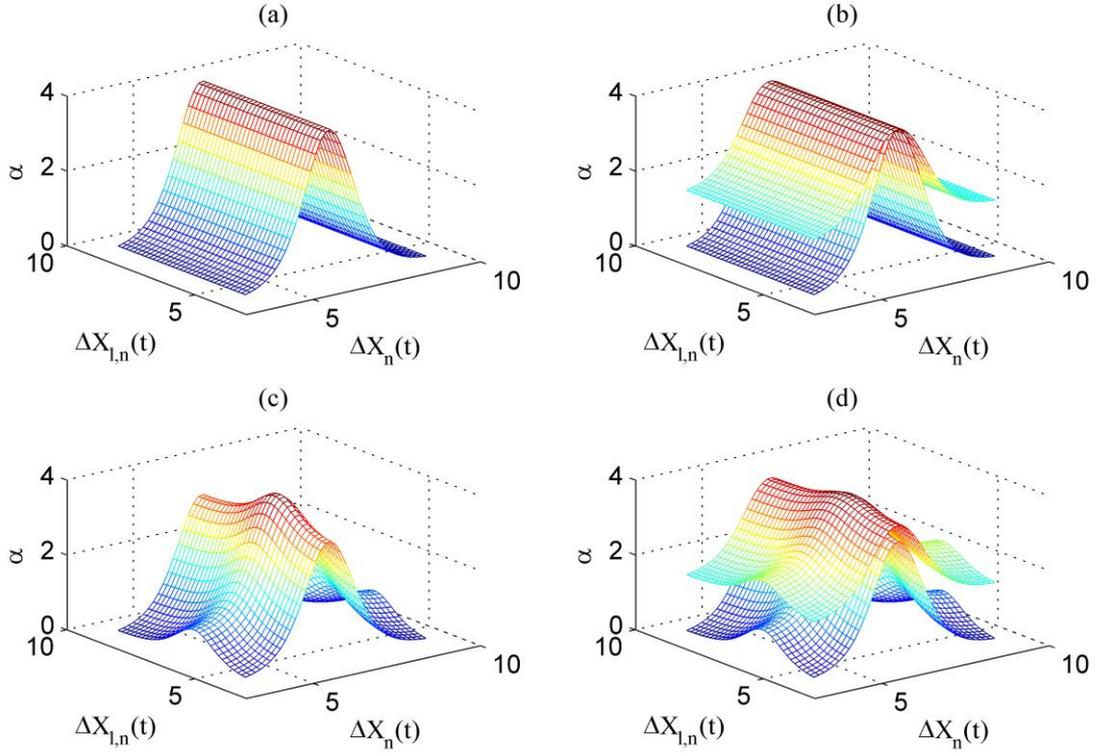

FIG.1. Headway-sensitivity space for (a) and (b): $p$=1, $q$=0; (c) and (d): $p$=0.8, $q$=0.2.

Then expand $z$ by the order of $ik$ at the point of $ik \approx 0$ as $z = z_1 ik + z_2(ik)^2 + \cdots$ and insert it into Eq. (8). The following terms can be obtained:

$$\begin{cases} z_1 = pV' + q\bar{V}' \\ z_2 = (\frac{1}{2} + \frac{\lambda_1 + \lambda_2}{a})(pV' + q\bar{V}') - \frac{(pV' + q\bar{V}')^2}{a} \end{cases}, \quad (9)$$

According to previous study, the vehicle system is stable when $z_2 > 0$, which is

$$a > 2(pV' + q\bar{V}') - 2(\lambda_1 + \lambda_2), \quad (10)$$

and the neutral stability condition has the following form:

$$a_c = 2(pV' + q\bar{V}') - 2(\lambda_1 + \lambda_2). \quad (11)$$

The stability surface is described in Fig. 1.

Parameters are set as $v_{\max} = 4m/s$, $h_c = 7m$, $p$=1 and $q$=0 in Fig. 1(a), while $p$=0.8 and $q$=0.2 in Fig. 1(c). As can be seen, the unstable region is smaller considering influence from adjacent lane [31].

## 4. Non-linear stability analysis

For the convenience of non-linear analysis, Eq. (5) is rewritten as

$$\begin{aligned} \frac{d^2 \Delta x_n(t)}{dt^2} &= \\ \alpha[pV(\Delta x_{n+1}(t)) &- pV(\Delta x_n(t)) + q\bar{V}(\Delta x_{l,n+1}(t)) \\ -q\bar{V}(\Delta x_{l,n}(t)) &- \frac{d\Delta x_n(t)}{dt}] + \lambda_1 \frac{d\Delta x_{n+1}(t)}{dt} \\ -\lambda_1 \frac{d\Delta x_n(t)}{dt} &+ \lambda_2 \frac{d\Delta x_{l,n+1}(t)}{dt} - \lambda_2 \frac{d\Delta x_{l,n}(t)}{dt}, \end{aligned} \quad (12)$$

MKdV equation is obtained in unstable

region around the critical point $(h_c, a_c)$, where $V'' = 0$. By the analysis method in [18-20], the long wave expansion is applied in this section. Two slow scales for space variable $n$ and time variable $t$ are introduced. We define slow variables $X$ and $T$ as

$$X = \varepsilon(n+bt), \quad T = \varepsilon^3 t, \quad (13)$$

where $b$ is a constant determined later and $\varepsilon = \sqrt{a_c/a - 1}$.

Headways for two lanes are set as

$$\begin{cases} \Delta x_n(t) = h_c + \varepsilon R(X,T) \\ \Delta x_{l,n}(t) = h_c + \varepsilon R(X,T) \end{cases}, \quad (14)$$

Expand Eq. (12) to the fifth order of $\varepsilon$, then gives

$$\varepsilon^2[ab - a(pV' + q\bar{V}')]\partial_X R + \varepsilon^3[b^2 - \frac{a(pV' + q\bar{V}')}{2}$$
$$-(\lambda_1 + \lambda_2)b]\partial_X^2 R + \varepsilon^4[a\partial_T R - \frac{a(pV' + q\bar{V}')}{6}\partial_X^3 R$$
$$-\frac{(\lambda_1 + \lambda_2)b}{2}\partial_X^3 R - \frac{a(pV''' + q\bar{V}''')}{6}\partial_X R^3] \quad (15)$$
$$+\varepsilon^5[(2b - \lambda_1 - \lambda_2)\partial_X \partial_T R - \frac{a(pV' + q\bar{V}')}{24}\partial_X^4 R$$
$$-\frac{(\lambda_1 + \lambda_2)b}{6}\partial_X^4 R - \frac{a(pV''' + q\bar{V}''')}{12}\partial_X^2 R^3] = 0.$$

It is noticed that the $\partial_X \partial_T R$ in the sixth order term of Eq. (15) can be eliminated by taking the derivative of $X$ in the fifth order term. Then insert $b = pV' + q\bar{V}'$ and $a_c/a - 1 = \varepsilon^2$ into Eq. (15), that is

$$[\partial_T R - m_1 \partial_X^3 R + m_2 \partial_X R^3] + \varepsilon[m_3 \partial_X^2 R + m_4 \partial_X^4 R + m_5 \partial_X^2 R^3] = 0, \quad (16)$$

where

$$m_1 = \frac{(a + 3\lambda_1 + 3\lambda_2)(pV' + q\bar{V}')}{6a},$$

$$m_2 = -\frac{(pV''' + q\bar{V}''')}{6},$$

$$m_3 = \frac{(pV' + q\bar{V}')}{2},$$

$$m_4 = \frac{4(2pV' + 2q\bar{V}' - \lambda_1 - \lambda_2)(a + 3\lambda_1 + 3\lambda_2)}{24a^2}$$
$$\times (pV' + q\bar{V}') - \frac{a + 4(\lambda_1 + \lambda_2)}{24a}(pV' + q\bar{V}'),$$

$$m_5 = \frac{2(2pV' + 2q\bar{V}' - \lambda_1 - \lambda_2) - a}{12a}(pV''' + q\bar{V}''').$$

In order to have standard MKdV equation, the following transformations are made

$$T = \frac{1}{m_1} T_m, \quad R = \sqrt{\frac{m_1}{m_2}} R_m. \quad (17)$$

Then Eq. (16) can be rewritten as

$$[\partial_{T_m} R_m - \partial_X^3 R_m + \partial_X R_m^3] + \frac{\varepsilon}{m_1}[m_3 \partial_X^2 R_m \quad (18)$$
$$+ m_4 \partial_X^4 R_m + \frac{m_1 m_5}{m_2} \partial_X^2 R_m^3] = 0.$$

Ignoring the $O(\varepsilon)$ term, we have MKdV equation with a kink-antikink soliton solution expressed as

$$R_{m0}(X, T_m) = \sqrt{B} \tanh[\sqrt{\frac{B}{2}}(X - BT_m)]. \quad (19)$$

To determine the value of amplitude $B$, the solvable condition is considered:

$$(R_{m0}, M[R_{m0}]) \equiv \int_{-\infty}^{\infty} R_{m0} M[R_{m0}] dX = 0, \quad (20)$$

where $M[R_{m0}]$ means the $O(\varepsilon)$ term in Eq. (18). By performing the integration of Eq. (20), the value of amplitude $B$ can be obtained:

$$B = \frac{5m_2 m_3}{2m_2 m_4 - 3m_1 m_5}. \quad (21)$$

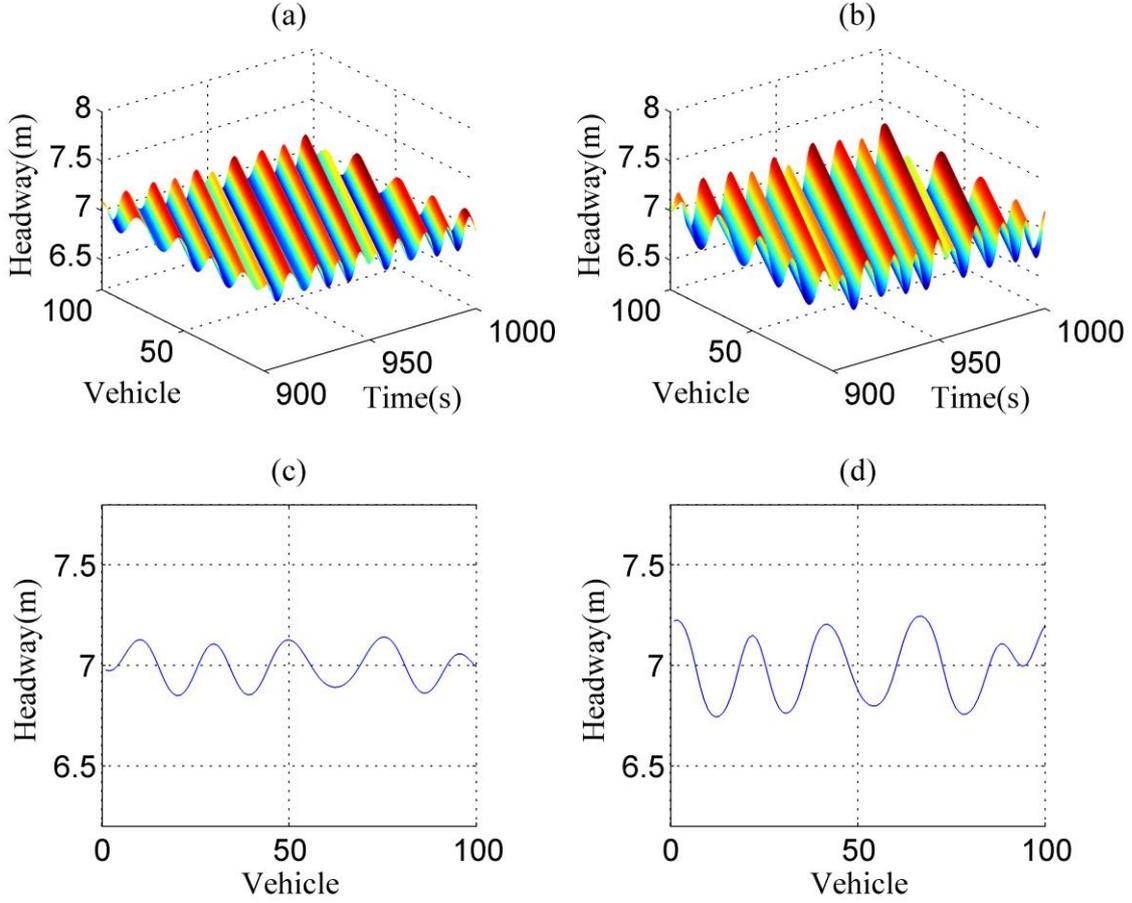

FIG.2. Space-time evolution of headway for lane1 in (a), lane2 in (b) and headway profile of traffic wave at $t$=950s for lane1 in (c), lane2 in (d). ($a$=2.85, $p$=1, $q$=0.)

The kink-antikink soliton solution of headway can be written as follows

$$\Delta x_n = h_c + \sqrt{(\frac{a_c}{a}-1)(\frac{5m_1 m_3}{2m_2 m_4 - 3m_1 m_5})}$$
$$\times \tanh\{\sqrt{(\frac{a_c}{a}-1)\frac{5m_2 m_3}{4m_2 m_4 - 6m_1 m_5}}[n+(pV'+q\bar{V}')t \quad (22)$$
$$-(\frac{a_c}{a}-1)\frac{5m_2 m_3 t}{2m_2 m_4 - 3m_1 m_5}]\}.$$

With the amplitude of Eq. (22), we have the coexisting surface in Fig. 1(b) and (d) based on (a) and (c), respectively. The space is divided into three regions: stable region above the coexisting surface, metastable region between coexisting surface and stability surface and unstable region below stability surface.

## 5. Numerical simulation

Consider a two lane system with 100 vehicles running on each lane under a periodic boundary condition without overtaking or lane changing [31]. The initial values are $v_{max}$=4m/s, $h_c$=7m and $d$=10 on both lanes. Perturbations are defined as follows:

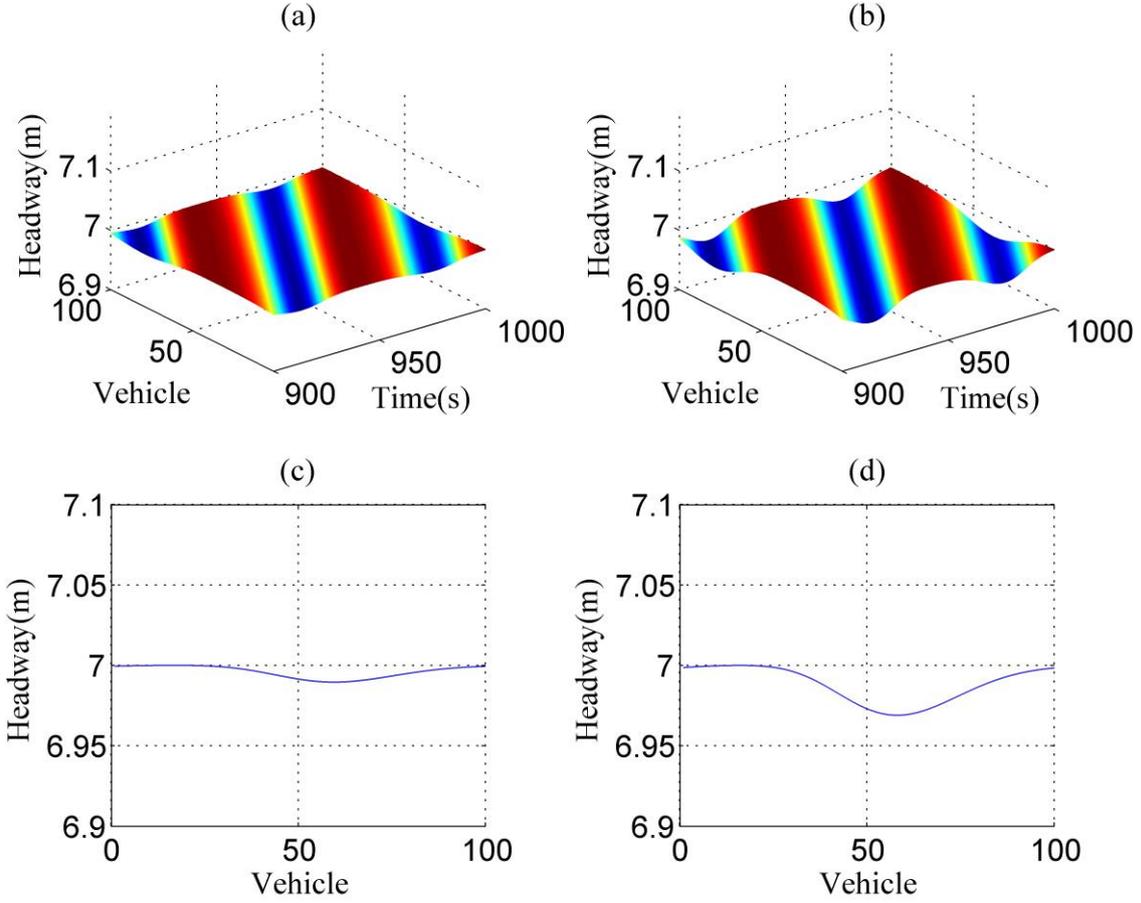

FIG.3. Space-time evolution of headway for lane1 in (a), lane2 in (b) and headway profile of traffic wave at $t$=950s for lane1 in (c), lane2 in (d). ($a$=2.85, $p$=0.8, $q$=0.2)

for lane1, $\begin{cases}\Delta x_n(0) = 7m - 0.1m & 45 < n < 50 \\ \Delta x_n(0) = 7m & others\end{cases}$;

for lane2, $\begin{cases}\Delta x_n(0) = 7m - 0.3m & 45 < n < 50 \\ \Delta x_n(0) = 7m & others\end{cases}$.

First let $a$=2.85, $p$=1, $q$=0, $\lambda_1$=0.2 and $\lambda_2$=0, which cannot satisfy the stable condition in Eq. (10). Figure 2 shows the space-time evolution of headway for lane1 in (a) and lane2 in (b) from 900s to 1000s. The headway profiles of traffic wave at $t$=950s are in (c) for lane1 and (d) for lane2. It is observed that the small perturbation propagates into traffic jam on both lanes. Furthermore, the traffic jam is more serious in lane2 because of the larger initial headway perturbation.

Then suppose $a$=2.85, $p$=0.8, $q$=0.2, $\lambda_1$=0.16 and $\lambda_2$=0.04. Figure 3 describes the space-time evolution of headway and the headway profiles of traffic wave at t=950s corresponding to Fig. 2. The initial perturbation decays after sufficient time on both lanes. Thus the consideration of lateral impact from adjacent lane can suppress traffic jam. The amplitude of traffic wave in lane2 is also larger.

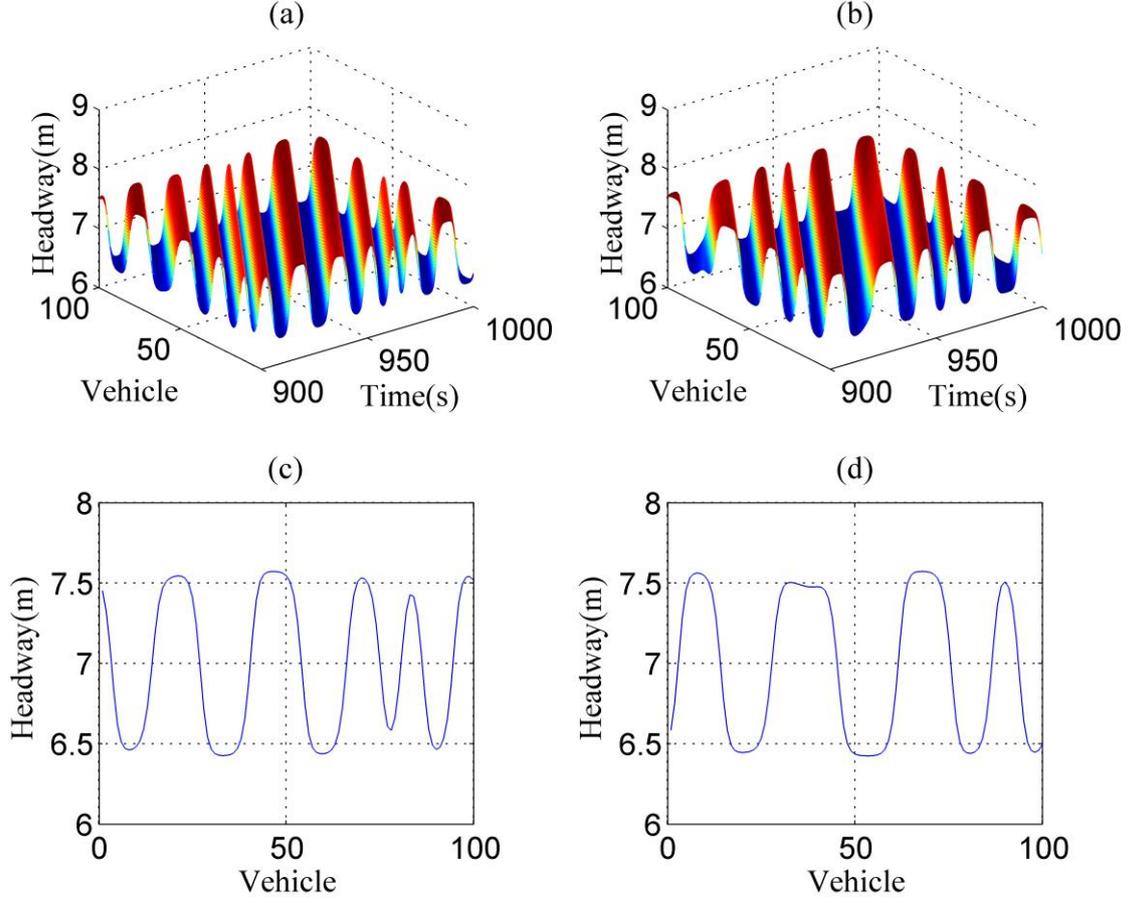

FIG.4. Space-time evolution of headway for lane1 in (a), lane2 in (b) and headway profile of traffic wave at $t$=950s for lane1 in (c), lane2 in (d). ($a$=2.2, $p$=0.8, $q$=0.2)

Finally when $a$=2.2, $p$=0.8, $q$=0.2, $\lambda_1$=0.16 and $\lambda_2$=0.04, on which situation the system is more unstable, we find serious kink-antikink waves on both lanes in Fig. 4. Unlike Fig. 2 or Fig. 3, there is no significant difference between the two lanes, because all the vehicles get influenced heavily by lateral impact. By simulation, the theoretical analysis of MKdV solution can be described.

**6. Conclusions**

In this paper, a new car-following model is proposed considering the lateral influence from adjacent lane. Both linear and non-linear stability analyses are carried out, from which MKdV equation is obtained. Numerical simulations show that the new model has good consistency with theoretical study. However, when satisfied the unstable condition, both lanes will have serious traffic jam of same level despite different initial headway perturbation. In conclusion, considering lateral impact indeed has influence on car-following behavior and can keep the vehicle system more stable.